\documentstyle[twoside]{article}

\catcode`\@=11
\long\def\@makefntext#1{
\protect\noindent \hbox to 3.2pt {\hskip-.9pt  
$^{{\eightrm\@thefnmark}}$\hfil}#1\hfill}		

\def\@makefnmark{\hbox to 0pt{$^{\@thefnmark}$\hss}}	
	
\def\ps@myheadings{\let\@mkboth\@gobbletwo
\def\@oddhead{\hbox{}
\rightmark\hfil\eightrm\thepage}   
\def\@oddfoot{}\def\@evenhead{\eightrm\thepage\hfil
\leftmark\hbox{}}\def\@evenfoot{}
\def\sectionmark##1{}\def\subsectionmark##1{}}

{\Large

\oddsidemargin=\evensidemargin
\addtolength{\oddsidemargin}{-30pt}
\addtolength{\evensidemargin}{-30pt}

\newcounter{sectionc}\newcounter{subsectionc}\newcounter{subsubsectionc}
\renewcommand{\section}[1] {\vspace{12pt}\addtocounter{sectionc}{1} 
\setcounter{subsectionc}{0}\setcounter{subsubsectionc}{0}\noindent 
	{\tenbf\thesectionc. #1}\par\vspace{5pt}}
\renewcommand{\subsection}[1] {\vspace{12pt}\addtocounter{subsectionc}{1} 
	\setcounter{subsubsectionc}{0}\noindent 
	{\bf\thesectionc.\thesubsectionc. {\kern1pt \bfit #1}}\par\vspace{5pt}}
\renewcommand{\subsubsection}[1] {\vspace{12pt}\addtocounter{subsubsectionc}{1}
	\noindent{\tenrm\thesectionc.\thesubsectionc.\thesubsubsectionc.
	{\kern1pt \tenit #1}}\par\vspace{5pt}}
\newcommand{\nonumsection}[1] {\vspace{12pt}\noindent{\tenbf #1}
	\par\vspace{5pt}}

\topsep=0in\parsep=0in\itemsep=0in
\parindent=15pt

\newcommand{\textlineskip}{\baselineskip=13pt}
\newcommand{\smalllineskip}{\baselineskip=10pt}

\def\eightcirc{
\begin{picture}(0,0)
\put(4.4,1.8){\circle{6.5}}
\end{picture}}
\def\eightcopyright{\eightcirc\kern2.7pt\hbox{\eightrm c}} 

\newcommand{\copyrightheading}[1]
	{\vspace*{-2.5cm}\smalllineskip{\flushleft
        {\footnotesize Nuovo Cimento B 114 (June 1999) 723-726 #1}\\
        {\footnotesize Los Alamos electronic archives: quant-ph/9809021
        v3=pv#1}\\
        {\footnotesize $\eightcopyright$\, H.C. Rosu
        }\\
	 }}

\def\abstracts#1#2#3{{
	\centering{\begin{minipage}{4.5in}\baselineskip=10pt\footnotesize
	\parindent=0pt #1\par 
	\parindent=15pt #2\par
	\parindent=15pt #3
	\end{minipage}}\par}} 

\renewenvironment{thebibliography}[1]
	{\frenchspacing
	 \ninerm\baselineskip=11pt
	 \begin{list}{\arabic{enumi}.}
        {\usecounter{enumi}\setlength{\parsep}{0pt}     
	 \setlength{\leftmargin 12.7pt}{\rightmargin 0pt} 
         \setlength{\itemsep}{0pt} \settowidth
	{\labelwidth}{#1.}\sloppy}}{\end{list}}

\newcounter{itemlistc}
\newcounter{romanlistc}
\newcounter{alphlistc}
\newcounter{arabiclistc}

\def\@citex[#1]#2{\if@filesw\immediate\write\@auxout
	{\string\citation{#2}}\fi
\def\@citea{}\@cite{\@for\@citeb:=#2\do
	{\@citea\def\@citea{,}\@ifundefined
	{b@\@citeb}{{\bf ?}\@warning
	{Citation `\@citeb' on page \thepage \space undefined}}
	{\csname b@\@citeb\endcsname}}}{#1}}

\newif\if@cghi
\def\cite{\@cghitrue\@ifnextchar [{\@tempswatrue
	\@citex}{\@tempswafalse\@citex[]}}
\def\citelow{\@cghifalse\@ifnextchar [{\@tempswatrue
	\@citex}{\@tempswafalse\@citex[]}}
\def\@cite#1#2{{$\null^{#1}$\if@tempswa\typeout
	{IJCGA warning: optional citation argument 
	ignored: `#2'} \fi}}

\def\@refcitex[#1]#2{\if@filesw\immediate\write\@auxout
	{\string\citation{#2}}\fi
\def\@citea{}\@refcite{\@for\@citeb:=#2\do
	{\@citea\def\@citea{, }\@ifundefined
	{b@\@citeb}{{\bf ?}\@warning
	{Citation `\@citeb' on page \thepage \space undefined}}
	\hbox{\csname b@\@citeb\endcsname}}}{#1}}

\def\@refcite#1#2{{#1\if@tempswa\typeout
        {IJCGA warning: optional citation argument
	ignored: `#2'} \fi}}

\def\refcite{\@ifnextchar[{\@tempswatrue
	\@refcitex}{\@tempswafalse\@refcitex[]}}

\def\pmb#1{\setbox0=\hbox{#1}
	\kern-.025em\copy0\kern-\wd0
	\kern.05em\copy0\kern-\wd0
	\kern-.025em\raise.0433em\box0}

\def\fnt#1#2{\footnotetext{\kern-.3em
	{$^{\mbox{\scriptsize #1}}$}{#2}}}

\headsep=15pt   
   
\font\tenrm=cmr10
\font\tenit=cmti10 
\font\tenbf=cmbx10
\font\bfit=cmbxti10 at 10pt
\font\ninerm=cmr9

\font\eightrm=cmr8

\textwidth=5truein
\textheight=7.8truein

\def\qed{\hbox{${\vcenter{\vbox{			
   \hrule height 0.4pt\hbox{\vrule width 0.4pt height 6pt
   \kern5pt\vrule width 0.4pt}\hrule height 0.4pt}}}$}}

\begin{document}



\normalsize\textlineskip
\thispagestyle{empty}
\setcounter{page}{1}

\copyrightheading{}			

\vspace*{0.88truein}

\centerline{\bf STRICTLY ISOSPECTRAL SUSY POTENTIALS
     }
\centerline{\bf  AND THE RICCATI SUPERPOSITION PRINCIPLE}
\vspace*{0.035truein}
\vspace*{0.37truein}
\centerline{\footnotesize H. ROSU}
\vspace*{0.015truein}
\centerline{\footnotesize\it Instituto de F\'{\i}sica,
Universidad de Guanajuato, Apdo Postal E-143, Le\'on, Gto, Mexico}
\baselineskip=10pt
\vspace*{10pt}

\vspace*{0.21truein}
\abstracts{{\bf Summary.} -
The connection of unbroken SUSY quantum mechanics in its strictly
isospectral form with the nonlinear Riccati superposition principle is pointed
out.
}{}{}


\textlineskip                  
\vspace*{12pt}                 

\vspace*{1pt}\textlineskip	
\bigskip


\bigskip

\noindent




\noindent
Generating quasi-isospectral
potentials by the methods of supersymmetric quantum mechanics (SUSYQM) 
is a simple and well-known mathematical technique [\refcite{Wi}].
Its relationships with intertwining, factorization, Darboux covariance,
and inverse scattering procedures have been recently reviewed by the author
[\refcite{Ro1}].
Also, the author, either in collaboration or alone [\refcite{Ro2}] dealt
extensively with a
SUSYQM procedure based on the general Riccati solution [\refcite{rsol}], that
may be called the double Darboux general Riccati method (henceforth DDGR).
For details the reader is directed to my previous works on DDGR and the
references therein.
The purpose of this note is to display the connection of the DDGR method
with the Riccati nonlinear superposition principle.

The DDGR procedure has been first used in physics by Mielnik for the quantum
harmonic
oscillator case [\refcite{KS}]. For simplicity reasons, I shall deal with
the case of unbroken SUSY, i.e.,
when one has at hand a true zero mode ${\rm u_0}$ of the initial
Schroedinger problem.
DDGR leads to
families of {\em strictly} isospectral potentials to the initial (bosonic)
one. Asking for the most general superpotential (i.e., the general
Riccati solution)
such that the fermionic potential fulfills
$\rm V_+(x)=  w_{g}^2 + \frac{d w_{g}}{dx}$, it is obvious that
one particular solution to this equation is $\rm w_p$, i.e., Witten's
superpotential.
Thus, one is led to consider the following
Riccati equation $\rm  w_{g}^2 + \frac{d w_{g}}{dx}=w^2_p +\frac{d w_p}{dx}$,
whose general solution can be written down as 
$\rm w_{g}(x)= w_p(x) + \frac{1}{v(x)}$, where $\rm v(x)$ is an unknown
function. Using this ansatz, one obtains for the function $\rm v(x)$ the 
following Bernoulli equation
\begin{equation}   \label {5}
\rm \frac{dv(x)}{dx} - 2 \, v(x)\, w_p(x) = 1,
\end{equation}
that has the solution
\begin{equation}  \label{6}
\rm v(x)= \frac{{\cal I}_0(x)+ \lambda}{u_{0}^{2}(x)},
\end{equation}
where $\rm {\cal I}_0(x)= \int^x \, u_0^2(y)\, dy $, and $\lambda$ is an
integration constant considered as the free parameter of the method.
Thus, $\rm w_{g}(x)$ can be written as follows
\begin{equation} \label{general}
\rm w_{g}(x;\lambda)
=\rm w_p(x) + \rm \frac{d}{dx} \Big[{\rm ln}
\left({\cal I}_0(x) + \lambda\right)\Big]
 =\rm w_p(x)+\sigma _{0}(\lambda) 
 =\rm - \frac{d}{dx} \Big[ ln \left(\frac{u_0(x)}{{\cal I}_0(x) +
\lambda}\right)
\Big],
\end{equation}
where in (2) and (3) the well-known relationship ($\rm D=d/dx$)
\begin{equation}  \label{particular}
\rm w_p(x)= -D \Big[ln \left(u_0(x)\right) \Big]=-\sigma _{0}
\end{equation}
has been used. The  
$\sigma _{0}$ notation for the logarithmic derivative is borrowed from
the book of Matveev and Salle [\refcite{MS}],
and the subscript has been appended to show that the fixed solution on
which the scheme is carried on is the Schroedinger ground state.
Finally, one gets the family of ``bosonic"
potentials 
\begin{eqnarray}
\rm  V_-(x;\lambda) &=& \rm w_{g}^2(x;\lambda) -
\frac{d w_{g}(x;\lambda)}{dx} \nonumber\\
&=& \rm V_-(x) - 2 D^2 \lbrack ln({\cal I}_0(x) + \lambda)
\rbrack \nonumber\\
&=& \rm V_-(x) -2\sigma _{0,x}(\lambda) \nonumber \\
&=& \rm V_-(x) - \frac{4 u_0(x) u_0^\prime (x)}{{\cal I}_0(x)
+ \lambda} 
+ \frac{2 u_0^4(x)}{({\cal I}_0(x) + \lambda)^2}.
\end{eqnarray}
All $\rm  V_-(x;\lambda)$ have the same supersymmetric partner potential
$\rm V_+(x)$ obtained by deleting the ground state.
 By inspection of Eqs.(\ref{particular}) and
(\ref{general}) we can obtain the ground state wave functions
for the potentials $\rm V_{-}(x;\lambda)$ as follows
\begin{equation} \label {9}
\rm u_0(x;\lambda)= f(\lambda)
\frac{u _0(x)}{{\cal I}_0(x) + \lambda},
\end{equation}
where $\rm f(\lambda)$ is a normalization factor of the form
$\rm f(\lambda)= \sqrt{\lambda(\lambda +1)}$. A connection with
other isospectral methods has been found, by noticing that
the limiting values -1 and 0 for the parameter $\lambda$ lead to the
Abraham-Moses procedure [\refcite {AbMo}], and Pursey's one [\refcite{Pu}],
respectively.
Here, we notice furthermore that by writing the parametric family in terms
of their unique ``fermionic" partner
\begin{eqnarray}
\rm  V_-(x;\lambda)
&=& \rm V_+(x) - 2 D^2 \Big[
ln\left(\frac{{\cal I}_0(x) + \lambda}{u_0}\right)
\Big] \nonumber\\
&=& \rm V_+(x) -2D^2 ln \left(\frac{1}{u_0(x;\lambda)}\right),
\end{eqnarray}
the inverse character of the parametric Darboux transformation is
manifest [\refcite{MS}].
Therefore, a two step interpretation is possible.
In the first step, one goes to the fermionic
partner system, and in the second one returns to a $\lambda$-distorted bosonic
system. This is why it can be called a double Darboux transformation, although
this may be somewhat confusing because of Adler's method [\refcite{adl}], which
is also a two step construction and is known by the same name. The strictly
isospectral DDGR method can be applied to any one-dimensional
system whose dynamics is dictated by a Schroedinger-like equation.
Moreover, one can employ combinations of any pairs of
Abraham-Moses procedure, Pursey's one, and the Darboux one.
However, only the DDGR leads to reflection
and transmission amplitudes identical to those of the original potential.

We now make use of the fact that Riccati equations are
differential equations admitting a
nonlinear superposition principle [\refcite{cmn}], i.e., it is possible to
write down the
general solution as a combination of particular solutions without using any
quadrature. This is due to the well-known property [\refcite{Dav}]
that given three particular solutions
one can obtain the general Riccati solution from the following ${\rm k}$
invariant
\begin{equation} \label {11}
\rm \frac{w-w_1}{w-w_2}:\frac{w_3-w_1}{w_3-w_2}=k,
\end{equation}
as the superposition
\begin{equation}  \label {12}
\rm w=\frac{kw_1(w_3-w_2)+w_2(w_1-w_3)}{k(w_3-w_2)+w_1-w_3}.
\end{equation}
In particular,
we have seen that within DDGR
the general Riccati solution can be written in the form
$\rm w(\lambda)=-\sigma _0+\sigma _0(\lambda)$. Considering three particular
solutions $\rm w(\lambda _{i})$, i=1,2,3,
one can write the above invariant as
\begin{equation}  \label {13}
\rm
\frac{\sigma _{0}(\lambda)-\sigma _{0}(\lambda _1)}{\sigma _{0}(\lambda)-
\sigma _{0}(\lambda _2)}:\frac{\sigma _{0}(\lambda _{3})-\sigma _{0}
(\lambda _1)}{\sigma _{0}(\lambda _3)-\sigma _{0}(\lambda _2)}=k,
\end{equation}
i.e., in terms of the $\lambda$-dependent logarithmic derivatives.
The nonlinear Riccati invariant k can be seen as a constraint on the Darboux
contributions to the parametric bosonic potentials or as a constraint on a
sequence
of their free parameters $\rm \lambda _{i}$ if written as follows
\begin{equation}  \label {bis}
\rm
\frac{\int \sigma _{0,x}(\lambda)-\int \sigma _{0,x}
(\lambda _1)}
{\int \sigma _{0,x}(\lambda)-
\int \sigma _{0,x}(\lambda _2)}:
\frac{\int \sigma _{0,x}(\lambda _3)-
\int \sigma _{0,x}
(\lambda _1)}{\int \sigma _{0,x}(\lambda _3)
-\int \sigma _{0,x}(\lambda _2)}=k~.
\end{equation}

It is worthwhile to notice that an even more general form of the
Riccati $\rm k$ invariant
can be written down if one do not stop at the one-parameter type of bosonic
Riccati solution. In another work [\refcite{mult}], I have provided a simple
multiple-parameter (iterative) generalization of Mielnik's construction. At
an arbitrary
$\rm i$-order of iteration, one can write the general Riccati
solution as $\rm w_{g}^{(i)}=w_{0}+
w_{\lambda _1}+...w_{\lambda _{i-1}}+w_{\lambda}$, where up to $\rm i-1$ the
$\lambda$ parameters are fixed and only the last parameter is a free one,
whereas $\rm w_{\lambda _{j}}=D\ln(\lambda _{j}+\int ^xF_{j-1})$.
The $F$'s are
integration factors of the type $\rm F_{j}=e^{-\int^{x}2w_{p}^{(j)}}$,
where $\rm w_{p}^{(j)}$ is the particular Riccati solution at order $\rm j$,
which is
of the same form as $\rm w_{g}^{(j)}$ but with no free parameter. Since all
these
Riccati solutions correspond to the same fermionic Schroedinger potential,
any triplet of multiple-parameter particular solutions that may be of
arbitrary hierarchical order $\rm j$ can enter in the Riccati $\rm k$
invariant and may be used to get the general solution at an arbitrary order.
Thus, one can write
\begin{equation}  \label {15}
\rm
\frac{\int \sigma _{0,x}^{(i)}(...,\lambda)-\int \sigma _{0,x}^{(j_1)}
(...,\lambda _{(j_1)})}
{\int \sigma _{0,x}^{(i)}(...,\lambda)-
\int \sigma _{0,x}^{(j_{2})}(...,\lambda _{j_2})}:
\frac{\int \sigma _{0,x}^{(j_3)}(...,\lambda _{j_3})-
\int \sigma _{0,x}^{(j_1)}
(...,\lambda _{j_1})}{\int \sigma _{0,x}^{(j_3)}(...,\lambda _{j_3})
-\int \sigma _{0,x}^{(j_2)}(...,\lambda _{j_2})}=k~.
\end{equation}

Finally, let us notice that instead of the zero mode solution one can employ
any other solution of the initial Hamiltonian $\rm H_{b}u(x)=\epsilon u(x)$,
where the so-called factorization energy $\epsilon$ should be less than
the ground state energy of $\rm H_{b}$, and an analytic continuation is
performed in the independent variable [\refcite{rot}] to turn $\rm u$ into a
nodeless function. In this way
no singularities with respect to the initial potential are introduced. For this
more general case the zero subindex in all the above equations is to be droped.

\bigskip
\bigskip

\noindent
This work was partially supported by the CONACyT Project 458100-5-25844E.
The author wishes to thank Prof. J.F. Cari\~nena for a stimulating
discussion.


\nonumsection{References}


}

\begin{thebibliography}{000}
\bibitem {Wi} WITTEN E., {\em Nucl. Phys. B}, {\bf 185} (1981) 513;
        COOPER F. and
        FREEDMAN B., {\em Ann. Phys. (N.Y)}, {\bf 146} (1983) 262;
        For a recent review,
        see COOPER F., KHARE A. and SUKHATME U., {\em Phys. Rep.},
        {\bf 251} (1995) 267.

\bibitem{Ro1} ROSU H.C., quant-ph/9809056, in Burgos
        Workshop {\em Symmetries in
        Quantum Mechanics and Quantum Optics} (Spain), to appear.

\bibitem{Ro2} See for example,
       BOYA L.J., ROSU H., SEGUI-SANTONJA A.J. and VILA F.J.,
       {\em Nuovo Cimento B}, {\bf 113} (1998) 409; ROSU H., {\em Nuovo
       Cimento B}, {\bf 113} (1998) 1431; BOYA L.J., ROSU H.C.,
       SEGUI-SANTONJA A.J., SOCORRO J., VILA F.J., {\em J. Phys. A},
       {\bf 31} (1998) 8835.

\bibitem{rsol} For the equivalence between the Riccati equation and the
       Schroedinger equation see, HALEY S.B., {\em Am. J. Phys.}, {\bf 65}
       (1997) 237; STAHLHOFEN A.A.,
       ``The Riccati equation: The common basis for
       SUSYQM and the factorization method", Duke Univ. report
       (unpublished, 1988); {\em Am. J. Phys.}, {\bf 67} (1999) 257.

\bibitem {KS} MIELNIK B., {\em J. Math. Phys.}, {\bf 25} (1984) 3387;
        NIETO M.M.,
        {\em Phys. Lett. B}, {\bf 145} (1984) 208; FERNANDEZ D.J.,
        {\em Lett. Math. Phys.}, {\bf 8} (1984) 337; KHARE A. and
        SUKHATME U., {\em J. Phys. A}, {\bf 22} (1989) 2847;
        {\em Phys. Rev. A}, {\bf 40} (1989) 6185;
        PAPPADEMOS J., SUKHATME U. and
        PAGNAMENTA A., {\em Phys. Rev. A}, {\bf 48} (1993) 3525; ROSU H.C.,
        {\em Phys. Rev. A}, {\bf 54} (1996) 2571; {\em Phys. Rev. E},
        {\bf 56} (1997) 2269.
\bibitem {MS} MATVEEV V.B. and SALLE M.A., {\em Darboux Transformations and
Solitons} (Springer, 1991).
\bibitem {AbMo} ABRAHAM P.B. and MOSES H.E.,
         {\em Phys. Rev. A}, {\bf 22} (1980) 1333.
\bibitem {Pu} PURSEY D.L., {\em Phys. Rev. D}, {\bf 33} (1986) 1048.
\bibitem{adl} ADLER V.E., {\em Theor. Math. Phys.}, {\bf 101} (1994) 1381;
BAGROV V.G. and SAMSONOV B.F., {\em Phys. Part. Nucl.}, {\bf 28} (1997) 374.
\bibitem {cmn} CARI\~NENA J.F., MARMO G. and
         NASARRE J., {\em Int. J. Mod. Phys. A}, {\bf 13} (1998) 3601.
\bibitem {Dav} DAVIS H.T., {\em Introduction to Nonlinear Differential and
Integral Equations} (Dover, 1962).
\bibitem{mult} ROSU H.C., quant-ph/9904007.
\bibitem{rot} See for example,
SAMSONOV B.F., {\em J. Math. Phys.}, {\bf 39} (1998) 967;
FERNANDEZ D.J., HUSSIN V., MIELNIK B., {\em Phys. Lett. A}, {\bf 244} (1998)
309.










\end{thebibliography}
\end{document}